\documentclass[aps,prl,twocolumn,a4paper,10pt,notitlepage,footinbib,superscriptaddress,showpacs]{revtex4-1}%
\usepackage[english]{babel}
\usepackage[T1]{fontenc}
\usepackage{endnotes}
\usepackage{amssymb,amsmath,amsfonts}
\usepackage{textcomp}
\usepackage{graphicx,color}
\usepackage[dvipsnames]{xcolor}
\usepackage[utf8x]{inputenc}
\usepackage{bm}

\begin{document}

\title{Supercoiling Enhances DNA Mobility by Reducing Threadings and Entanglements} 

\author{Jan Smrek}
\affiliation{Faculty of Physics, University of Vienna, Boltzmanngasse 5, A-1090 Vienna, Austria}
\author{Davide Michieletto}
\affiliation{School of Physics and Astronomy, University of Edinburgh, Peter Guthrie Tait Road, Edinburgh, EH9 3FD, UK}
\affiliation{MRC Human Genetics Unit, Institute of Genetics and Molecular Medicine, University of Edinburgh, Edinburgh EH4 2XU, UK}

\begin{abstract}
DNA is increasingly employed for bio and nanotechnology thanks to its exquisite versatility and designability. Most of its use is limited to linearised and torsionally relaxed DNA but non-trivial architectures and torsionally constrained -- or supercoiled -- DNA plasmids are largely neglected; this is partly due to the limited understanding of how supercoiling affects the rheology of entangled DNA. To address this open question we perform large scale Molecular Dynamics (MD) simulations of entangled solutions of DNA plasmids modelled as twistable chains. We discover that, contrarily to what generally assumed in the literature, larger supercoiling increases the average size of plasmids in the entangled regime. At the same time, we discover that this is accompanied by an unexpected increase of diffusivity. We explain our findings as due to a decrease in inter-plasmids threadings and entanglements.
\end{abstract}

\maketitle 
	
	The deoxyribonucleic acid (DNA) is not only the central molecule of life but it is now increasingly employed for bio-compatible and responsive materials -- such as DNA hydrogels~\cite{Um2006} and origami~\cite{Rothemund2006} -- with applications in medicine and nanotechnology~\cite{Seeman2017}. One feature that renders DNA a unique polymer is its ability to encode information, and this is now extensively leveraged to make complex structures~\cite{Mao1997,Castro2011,Seeman2017} and even self-replicating materials~\cite{Leunissen2009}; another feature that distinguishes DNA from other synthetic polymers is its unique geometry, i.e. that of a (right-handed) helix with a well-defined pitch, which entails that DNA can display both bending and torsional stiffness~\cite{Bates2005}. Oppositely to the information-encoding feature of DNA, its geometrical features are far less exploited to create synthetic materials. In fact, DNA is at present largely employed in its simplest geometrical form, i.e. that of a linear or nicked circular (torsionally unconstrained) molecule~\cite{Robertson2006,Robertson2007,Teixeira2007,Fitzpatrick2018, Michieletto2019a}. In spite of this, most of the naturally occurring DNA is under torsional and topological constraints, either because circular and non-nicked (as in bacteria) or because of the binding of proteins that restrict the relative rotation of base-pairs in eukaryotes~\cite{Racko2018,Benedetti2014,Gilbert2014,Naughton2013}. 
The torsional stress stored in a closed DNA molecule cannot be relaxed but only re-arranged or converted into bending in order to minimise the overall conformational free energy~\cite{Irobalieva2015,Fosado2017,Vologodskii2016}. This entails that supercoiling -- the linking deficit between sister DNA strands with respect to their relaxed state -- can potentially carry conformational information~\cite{Sutthibutpong2016} which can even regulate gene transcription~\cite{Ding2014}. Here, we propose that supercoiling may also be leveraged to tune the dynamics of DNA plasmids in solution, thus potentially allow for a fine control over the rheology of DNA-based complex fluids in a way that is orthogonal to varying DNA length~\cite{Laib2006}, concentration~\cite{Zhu2008} or architecture~\cite{Robertson2007,Rosa2020}. 

Entangled solutions of DNA plasmids are not only interesting because of potential applications in bio and nanotechnology but also because they are a proxy to study fundamental questions on the physics of ring polymers~\cite{Robertson2007,Soh2019,Krajina2018}. 
To characterise the effect of DNA supercoiling in entangled solutions, here we perform large scale Molecular Dynamics simulations of entangled DNA plasmids (Fig.~\ref{fig:model}A-C), modelled as coarse-grained twistable chains~\cite{Brackley2014}. We discover that while isolated DNA plasmids typically display a collapse with increasing levels of supercoiling (estimated via simulations~\cite{Krajina2016} or gel electrophoresis~\cite{Cebrian2014}), here we show that entangled DNA plasmids typically increase their average size for increasing values of supercoiling. 
Importantly, we further discover that in spite of this, larger supercoiling is accompanied by an \emph{enhanced} mobility of the plasmids. This finding is counter-intuitive and in marked contrast with standard polymer systems~\cite{Doi1988} whereby larger polymer sizes correlate with slower diffusion. 
Finally, we quantify the abundance of threadings and entanglements and discover that larger supercoiling decreases both these topological constraints, hence explaining the enhanced mobility. Given the lack of detailed studies on the dynamics of entangled supercoiled plasmids and their current technical feasibility~\cite{Peddireddy2019}, we argue that achieving a better understanding of these systems will not only suggest new strategies to realise the next-generation of biomimetic DNA-based materials but also shed light into the dynamics of DNA \emph{in vivo}. 
\vspace*{-0.2 cm }

\section*{Results and Discussion}

\subsection*{Model}

\begin{figure}[t!]
	\centering
	\includegraphics[width=0.45\textwidth]{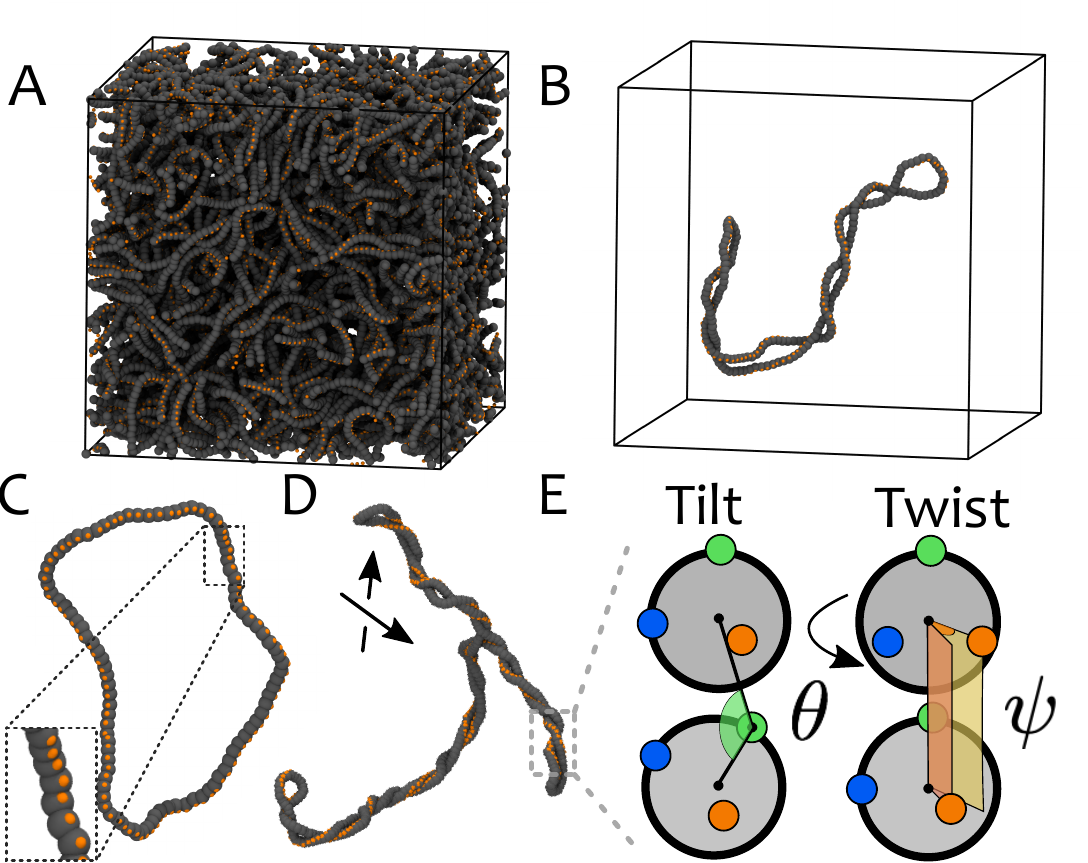}
			\vspace{-0.3 cm}
	\caption{\textbf{A} Snapshot of simulation of entangled plasmids with length $L=200\sigma_b \simeq 1.47 kbp \simeq 0.5 \mu m$ and $\sigma = 0.04$. \textbf{B} A single plasmid taken from \textbf{A}.  \textbf{C,D} Snapshots of other plasmids in solutions with (\textbf{C}) $\sigma = 0$, $L = 0.25 \mu m$ and (\textbf{D}) $\sigma = -0.06$, $L \simeq 1 \mu m$. Backbone beads are shown in grey, one set of patches are shown in orange. The other patches are not shown for clarity. \textbf{E} Sketch of tilt $\theta$ and tilt $\psi_1$ between consecutive beads (an equivalent $\psi_2$ is set between blue patches, not shown). The tilt angle $\theta$ is subject to a stiff potential with equilibrium $\theta_0=0$ to maintain the frame co-planar and aligned with the backbone.}
	\label{fig:model}
			\vspace{-0.6 cm}
\end{figure}

DNA is represented as a twistable elastic chain~\cite{Marko1994,Brackley2014} made of beads of size $\sigma_b=2.5$ nm$= 7.35$ bp connected by finitely-extensible springs and interacting via a purely repulsive Lennard-Jones potential to avoid spontaneous chain-crossing~\cite{Kremer1990}. In addition to these potentials, a bending stiffness of $l_p=50$ nm~\cite{Bates2005} is set via a Kratky-Porod term and two torsional springs (dihedrals) constrain the relative rotation of consecutive beads, $\psi$, at a user-defined value $\psi_0$. The torsional angle between consecutive beads $\psi$ is determined by decorating each bead with three patches which provides a reference frame running along the DNA backbone. We finally impose a stiff harmonic spring to constrain the tilt angle $\theta=\pi$ so to align the frame with the backbone, i.e. along its local tangent (see Fig.~\ref{fig:model}D). The simulations are performed at fixed monomer density $\rho \sigma_b^3 = 0.08$ and by evolving the equation of motion for the beads coupled to a heat bath in LAMMPS~\cite{Plimpton1995} (see Methods). 

The user-defined angle $\psi_0$ directly determines the thermodynamically preferred pitch of the twisted ribbon as $p = 2\pi/\psi_0$ and, in turn, this fixes the preferred linking number to $Lk = M/p$, where $M$ is the number of beads in the plasmid. The twist is enforced by a harmonic potential with stiffness $\kappa_t = 50\sigma_b = 125$ nm comparable with the torsional persistence length of DNA.
In this model, the relaxed linking number is $Lk_0 = 0$ and so the degree of supercoiling $\sigma \equiv  \Delta Lk/M = Lk/M = 1/p$. This is set by initialising the patchy-polymer as a flat ribbon and by subsequently imposing the angle $\psi_0$ in order to achieve the desired $\sigma$ (which may be zero, if $\psi_0=0$  or $p=\infty$). It should be noted that we also will consider relaxed plasmids in which the torsional stiffness is set to $\kappa_t=0$ mimicking fully nicked plasmids.
Finally, it should be recalled that for non-nicked circular DNA, the exchange of local torsion (twist $Tw$) into bending (writhe $Wr$) must obey the White-Fuller-C\u{a}lug\u{a}reanu (WFC)~\cite{Dennis2005,Fuller1971,White1987} theorem, i.e. $Lk = Tw + Wr$, thus conserving the linking number $Lk$ (and thus the supercoiling $\sigma=\Delta Lk/M$) between the two DNA single strands (Fig.~\ref{fig:model}B-D). Notice that our polymer model is symmetric with respect to supercoiling and therefore we will refer to $\sigma$ without specifying the sign. 


\vspace{-0.4 cm}
\subsection*{Supercoiling Increases the Average Size of DNA Plasmids in Entangled Conditions}

\begin{figure*}[t]
	\centering
	\includegraphics[width=0.95\textwidth]{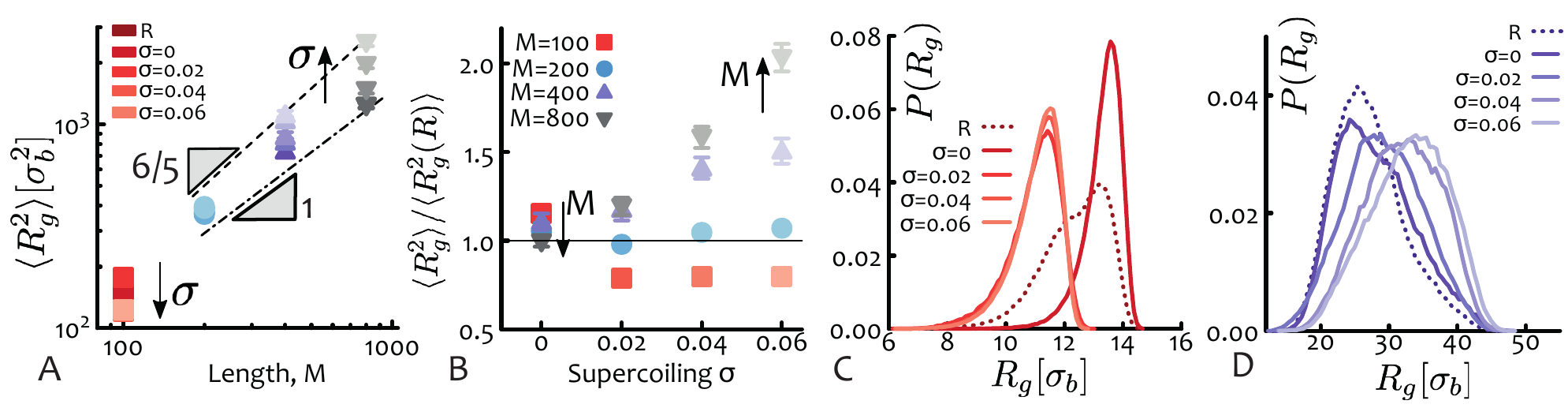}
			\vspace{-0.3 cm}
	\caption{\textbf{Supercoiling Increases Plasmids Size in Entangled Conditions.} \textbf{A-B} Radius of gyrations $R_g$ plotted against (\textbf{A}) contour length $M$ and (\textbf{B}) supercoiling $\sigma$. Notice that for short lengths $M=100$, increasing $\sigma$ induces a collapse of the plasmids whereas for longer lengths it drives swelling. The scaling of $R_g$ as a function of plasmid length is compatible with those of  ideal and self-avoiding walks for relaxed and highly supercoiled plasmids, respectively. \textbf{C} The distribution of $R_g$ for $M=100$ is weakly bimodal showing that plasmids can be in either an ``open'' or a ``collapsed'' state. Setting a supercoiling $\sigma=0$ stabilises the open state whereas $\sigma>0$ induces writhing and collapse. \textbf{D} For longer plasmids ($M=400$) larger supercoiling $\sigma$ broadens the distribution and drives enlarges the average size. The unit of length is $\sigma_b=2.5$ nm.  }
	\label{fig:RG}
			\vspace{-0.3 cm}
\end{figure*}

The conformational properties of polymers in solution are typically studied in terms of the gyration tensor 
\begin{equation}
R_T^{\alpha \beta} = \dfrac{1}{2 M^2} \sum_{i,j=1}^M \left(r^{\alpha}_i - r_j^{\alpha}\right) \left(r^{\beta}_i - r_j^{\beta}\right)
\end{equation} 
where $r_i^\alpha$ denotes the coordinate $\alpha$ of the position of bead $i$. The (square) radius of gyration is then defined as the trace, $R_g^2 \equiv \rm{Tr}[R_T]$. Interestingly, we find that the time and ensemble average of $R_g^2$ scales as $\langle R_g^2 \rangle \sim L^{2 \nu}$, with the metric exponent $\nu=1/2$ and $\nu=3/5$ for relaxed and highly supercoiled plasmids, respectively (for $M \geq 200$, see Fig.~\ref{fig:RG}A and Fig.~S1 in SI). These two regimes are compatible with the ones for ideal ($\nu=1/2$) and self-avoiding ($\nu \simeq 0.588$) random walks; this finding suggests that relaxed plasmids in entangled solutions ($\nu=1/2$) assume conformations similar to the ones of standard flexible, albeit short, ring polymers~\cite{Halverson2011statics}. At the same time, for larger values of $\sigma$, the self-interactions driven by writhing (see Fig.~\ref{fig:model}B,C) are no longer screened by the neighbours and we thus observe a larger metric exponent $\nu$ compatible with a self-avoiding walk. 

The effect of supercoiling on the average size of plasmids can be better appreciated in Fig.~\ref{fig:RG}B where we show the (squared) radius of gyration rescaled by its value for relaxed plasmids and plotted against supercoiling. One can readily notice that, importantly, for long plasmids (e.g. $M \geq 400 \simeq 3$ kb) the greater the supercoiling the \emph{monotonically} larger their typical size. We highlight that this behaviour is highly counter-intuitive as one expects that supercoiling induces the compaction of a polymer. For instance, supercoiled plasmids travel faster than their relaxed counterparts in gel electrophoresis~\cite{Cebrian2014}. Additionally, supercoiling is often associated with the packaging of the bacterial genome~\cite{Sinden1981,Wu2019} and with organisation into topological domains in eukaryotes~\cite{Gilbert2014a,Benedetti2014,Benedetti2017}. Interestingly, the observed monotonic increase of $R_g$  with supercoiling is in marked contrast with the overall shrinking seen in dilute conditions~\cite{Krajina2016} (which is also reproduced by our model in dilute conditions, see Fig.~S1 in SI). We argue that this stark difference is due to inter-chain effects as it is mostly evident for long chains but remarkably absent for short ones (Fig.~\ref{fig:RG}B). 

For short plasmids ($M=100 \simeq 730 bp$) we observe an interesting exception to the behaviour described above whereby the typical size is non-monotonic for increasing supercoiling levels. More specifically, for $\sigma=0$ we find that the conformations are typically larger than the relaxed ones, but they suddenly become more collapsed for $\sigma>0$ (Fig.~\ref{fig:RG}B). To investigate this behaviour further we examined the distributions of  radius of gyration and noticed that relaxed short plasmids display a weakly bimodal distribution that is not found in larger plasmids (Fig.~\ref{fig:RG}C,D). This bimodal distribution reflects the fact that short relaxed plasmids can be found in two typical conformation states: either open (large $R_g$) or more collapsed (small $R_g$). The switching between these two typical conformations is instead forbidden when a certain supercoiling is imposed as all allowed conformations must satisfy the WFC topological conservation law. More specifically, zero supercoiling ($\sigma=0$) hinders the writhing of the plasmid and, given their short length ($L/l_p=5$), it would be energetically too costly to writhe multiple times with opposite sign to achieve a null global writhe. This entails that short plasmids are locked into open, not self-entangled conformations. On the contrary, for $\sigma>0$, the imposed writhing induces a conformational collapse, akin to a sharp buckling transition~\cite{Ott2020}. 

We note that the stable open state at $\sigma=0$ for short plasmids is akin to the one computationally observed in dense solutions of semiflexible rings~\cite{Bernabei2013}. These systems are expected to give rise to exotic columnar phases which would be thus intriguing to investigate in the context of dense solutions of short null-supercoiled plasmids. 

We finally stress once more that the monotonic increase observed for long plasmids of their typical size with supercoiling is neither expected nor trivial and is in marked contrast with the overall shrinking behaviour found in the literature for long dilute supercoiled plasmids~\cite{Krajina2016}. Since the monomer concentration is constant for all the system studied, one would na\"ively expect that the effective $c/c^*$ is larger (the critical entanglement concentration scales as $c^* \sim M/R_g^3$). In turn, this implies more inter-chain entanglements which add up to the already more numerous intra-chain entanglements due to the increase of self-crossings (writhing). As a consequence, we also expect highly supercoiled long plasmids to have reduced mobility with respect to their relaxed counterparts. In summary, based on the results in Fig.~\ref{fig:RG} and simple arguments, we expect that supercoiling hinders the diffusion of plasmids. 
\vspace*{-0.2cm}

\subsection*{Supercoiling Enhances DNA Mobility}

\begin{figure}[t]
	\centering
	\includegraphics[width=0.47\textwidth]{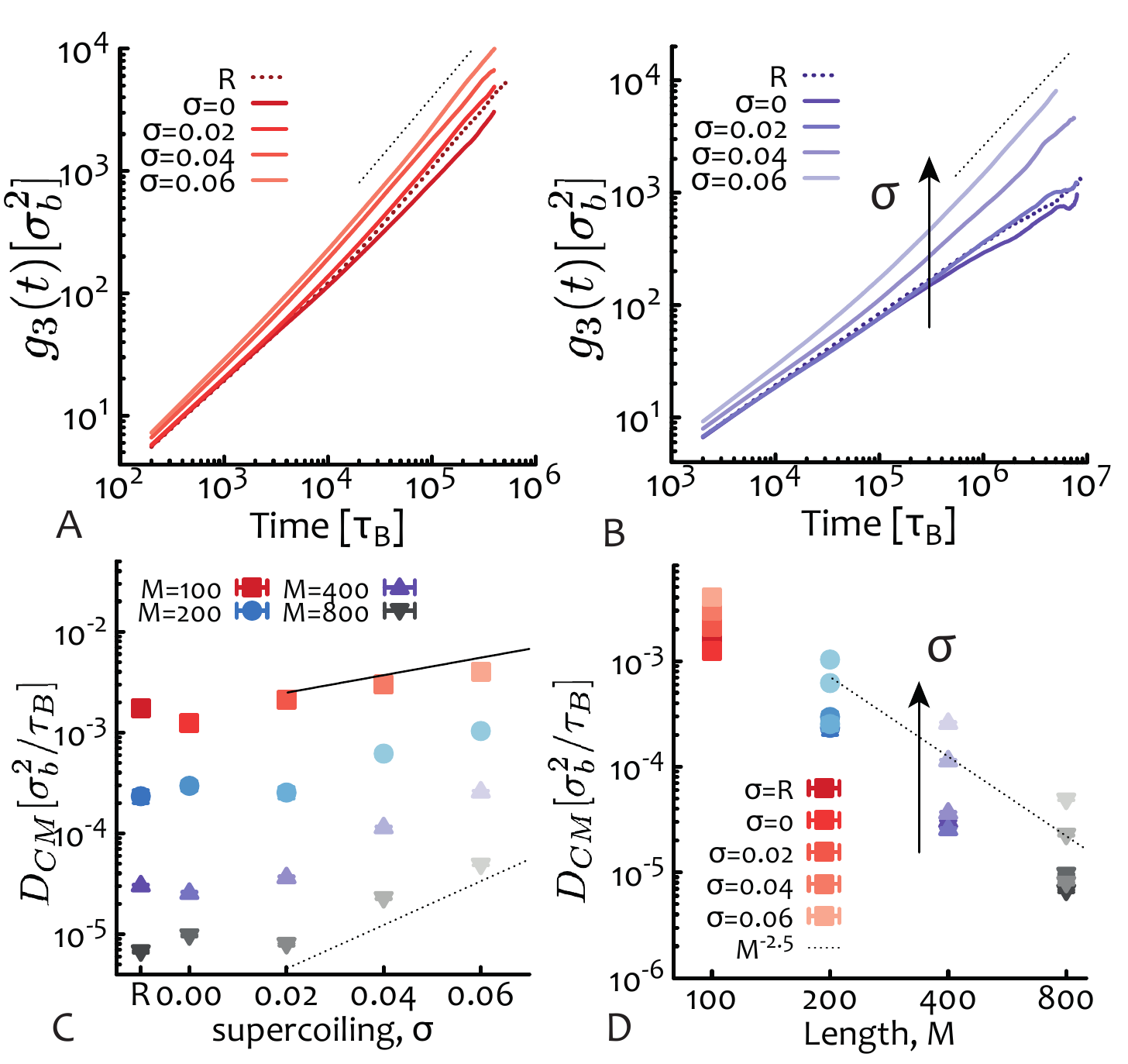}
			\vspace{-0.4 cm}
	\caption{ \textbf{Supercoiling Enhances Plasmid Mobility. }\textbf{A-B} TAMSD of the plasmids for (\textbf{A}) $M=100 \simeq 730$ bp) and (\textbf{B}) $M=400 \simeq 3$ kbp). Dashed lines are linear functions of lagtime as a guide for the eye.  \textbf{C-D} Diffusion coefficient of the centre of mass  $D_{CM}=\lim_{t\to \infty} g_3(t)/6t$ against (\textbf{C}) supercoiling $\sigma$ and (\textbf{D}) length $M$. In \textbf{C} exponentials $\sim \exp{\left(\sigma/0.05\right)}$ (solid) and $\sim \exp{\left(\sigma/0.02\right)}$ (dashed) are drawn as guide for the eye. In \textbf{D} a power law $\sim M^{-2.5}$ is fitted through the data for $\sigma=0.04$. Error bars are comparable to symbol size. R = ``relaxed''.}
	\label{fig:TAMSD}
			\vspace{-0.6 cm}
\end{figure}

We study the dynamics of the entangled plasmids at different levels of supercoiling by computing the time- and ensemble-averaged mean squared displacement (TAMSD) of the centre of mass (CM) of the plasmids as $g_3(t)=\langle \bm{r}_{CM,i}(t+t_0) - \bm{r}_{CM,i}(t_0) \rangle_{i,t_0}$ (other $g_i$ quantities are reported in Fig.~S4 in SI). Curves for $g_3$ are shown in Fig.~\ref{fig:TAMSD}A,B for different values of plasmid supercoiling and length. Strikingly, we find that higher values of $\sigma$ yield faster mobility especially for longer plasmids. We recall that this is at first unexpected given the monotonically larger size discussed in the previous section.

\begin{figure*}[t]
	\centering
	\includegraphics[width=0.95\textwidth]{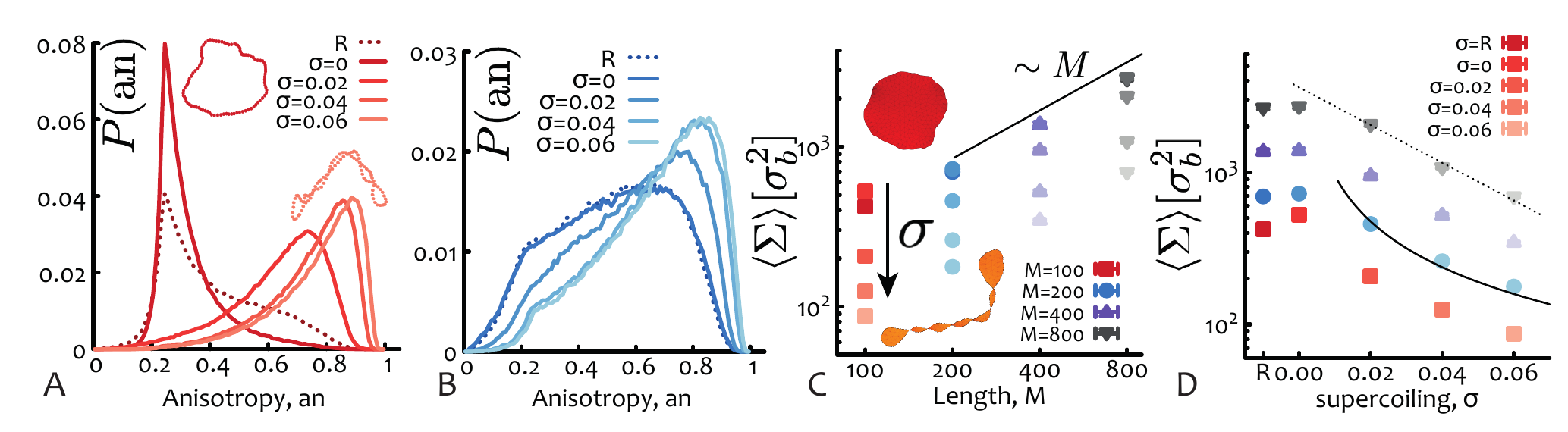}
	\vspace{-0.4 cm}
	\caption{\textbf{Supercoiling Induces Buckling in Short Plasmids and Reduces the Threadable Area.} \textbf{A} The anisotropy shape descriptor (an, see text) for short plasmids $M=100$ displays a sharp buckling transition between an open and roughly symmetric state for $\sigma=0$ and a collapsed and anisotropic one for $\sigma>0$. In inset, two examples of conformations are shown. \textbf{B} For longer plasmids ($M=200$) supercoiling shifts the anistropy to larger values indicating a smoother transition to more prolate conformations. \textbf{C} Scaling of the average minimal surface size $\langle \Sigma \rangle$ as a function of plasmids length (solid line shows the linear scaling). In inset, two examples of surfaces for $M=100$ are shown. \textbf{D} The size of the minimal surface area is monotonically decreased by supercoiling (with the exception of short $M \leq 200$ plasmids). The solid and dashed lines scale as $1/\sigma$ and $e^{-\sigma/0.035}$, respectively, and are drawn as a guide for the eye. R stands for ``relaxed''. The unit of length is $\sigma_b=2.5$ nm. The error bars, typically smaller than the symbol size represent the error of the mean area.}
		\vspace{-0.3 cm}
	\label{fig:Shape}
\end{figure*}

The diffusion coefficient of the centre of mass computed as  $D_{CM} = \lim_{t\to \infty} g_3(t)/6t$ allows us to more precisely quantify how the mobility of the plasmids changes as a function of length and supercoiling. 
In particular, we find that while $D_{CM}$ attains a plateau at small $\sigma$, at larger supercoiling it increases exponentially (see Fig.~\ref{fig:TAMSD}C) albeit more simulations are needed to confirm this conjecture (see below for an argument supporting the exponentially faster mobility). Additionally, we find that the diffusion coefficient as a function of plasmid length scales as $D_{CM} \sim M^{-2.5}$, compatible with the scaling of flexible and much longer ring polymers~\cite{Halverson2011dynamics} (Fig.~\ref{fig:TAMSD}D). [Note that the solutions with $M=800$ are not displaying a freely diffusive behaviour in spite of the fact that we ran them for more than $10^7$ Brownian times (see Tab.~S1 in SI); in turn, $D_{CM}$  is overestimated as its calculation assumes free diffusion. In spite of this, values of $D_{CM}$ for $M=800$ nicely follow the general trend of the other datasets (see Fig.~\ref{fig:TAMSD}C,D).]
 
It should be highlighted that the agreement of the dynamical exponent found here with the one for longer and more flexible rings is unexpected, as for shorter ring polymers (smaller $L/l_p$ and comparable with the rings studied here) the dynamical exponent was shown to be smaller (about $1.5$~\cite{Halverson2011dynamics}). We speculate that this more severe dependence on plasmid length may be due to the fact that DNA has a large persistence length ($l_p=20\sigma_b\simeq 50$ nm) which can thus reduce the effective entanglement length~\cite{Rosa2013}. In agreement with this argument, entangled solutions of semi-flexible and stiff ring polymers vitrify in certain concentration regimes~\cite{Slimani2014} implying that large stiffness may exacerbate topological constraints. In particular, the observed jamming was conjectured to be related to threading~\cite{Michieletto2016pnas,Michieletto2017prl,Smrek2016,Smrek2020}, allowed by the swelling of the rings' conformations~\cite{Slimani2014}.  
Motivated by this relationship between ring conformation and dynamics we now investigate in more detail the conformation properties of plasmids using shape descriptors derived from the gyration tensor.
\vspace{-0.2 cm}


\subsection*{Supercoiling Induces a Buckling Transition in Short Plasmids}

The consequence of writhing on the plasmids conformations is not captured by $R_g$ alone~\cite{Rawdon2008,Benedetti2015a}. Instead, it is informative to study shape descriptors which can be computed via the eigenvalues of the gyration tensor $R_T$ (which we denote as $a,b,c$, with $a>b>c$ and $R_g^2=a+b+c$). Typical shape descriptors are the asphericity~\cite{Rawdon2008,Benedetti2015a,Rosa2011}  
$a = ((a - b)^2 + (a-c)^2 + (b-c)^2)/2R_g^4$ 
which quantifies the deviation from a perfectly spherical arrangement and the nature of asphericity quantified by either the prolateness (see Fig.~S2 in SI) or the anisotropy $an=3(a^2+b^2+c^2)/(2 R_g^4)-1/2$ (shown in Fig.~\ref{fig:Shape}A,B). These shape descriptors reveal that for $M=100$ and $\sigma=0$, plasmids are stabilised in an open, highly symmetric and oblate (M\&M's) state. Furthermore, they reveal that these short plasmids undergo a buckling transition to a closed, asymmetric and prolate (rugby ball) shape for $\sigma > 0$. The sharp first-order-like buckling transition (see Fig.~\ref{fig:Shape}A and SI) is weakened for larger contour lengths (see Fig.~\ref{fig:Shape}B), as self-writhing is energetically allowed even for $\sigma=0$ (negative and positive self-crossings must cancel each other to satisfy the WFC conservation law).  At the same time, both short and long plasmids display a general increase in asphericity, prolateness and anisotropy with increasing supercoiling, strongly suggesting that the plasmids assume elongated and double-folded conformations (see Fig.~S2 in SI). \vspace{-0.2 cm}

\subsection*{Supercoiling Decreases the Spanning Minimal Surface}

It is natural to associate the open oblate and closed prolate conformations assumed by DNA plasmids to a larger and a smaller (minimal) area spanned by the contour, respectively~\cite{Lang2013}. The size of this area may be relevant for the dynamics because it could be ``threaded'' by neighbouring plasmids hence hindering the dynamics~\cite{Michieletto2016pnas,Smrek2016,Gomez2020}. To quantify this in more detail we calculated the minimal surface using the algorithm 
used in Ref.~\cite{Lang2013,Smrek2016,Smrek2019a} for relaxed ring polymers. We found that the minimal area grows linearly with the plasmids' contour, as expected~\cite{Smrek2016} (Fig.~\ref{fig:Shape}C). Importantly, we also observed that it overall decreased with supercoiling with the notable exception of short $M \leq 200$ plasmids, for which there is a small increase for $\sigma=0$ with respect to the relaxed case, again confirming the topological locking in open conformations (Fig.~\ref{fig:RG}A).

A crude way to estimate the decrease in ``threadable'' area of a plasmid is via recursive bisections of a perfect circle into several connected smaller circles joined at a vertex mimicking writhe-induced self-crossing. Each times a circle is split into two smaller ones the new radii are $R^\prime \simeq R/2$ and thus $n$ circles (with $n-1$ self-crossings) have radii $R^\prime = R/n$ yielding an overall spanning surface $\simeq n \pi (R/n)^2 \sim 1/n \sim 1/\sigma$.  This crude estimation is in good agreement with the behaviour of the minimal surface albeit we cannot rule out other functional forms (for instance exponential, see Fig.~\ref{fig:Shape}D). [Note that the so-called magnetic moment and radius~\cite{Schram2019} give similar results albeit different scaling (see Fig.~S3 in SI)]. 
\vspace{-0.2 cm}

\subsection*{Supercoiling Reduces Threadings}

\begin{figure*}[t]
	\centering
	\includegraphics[width=0.7\textwidth]{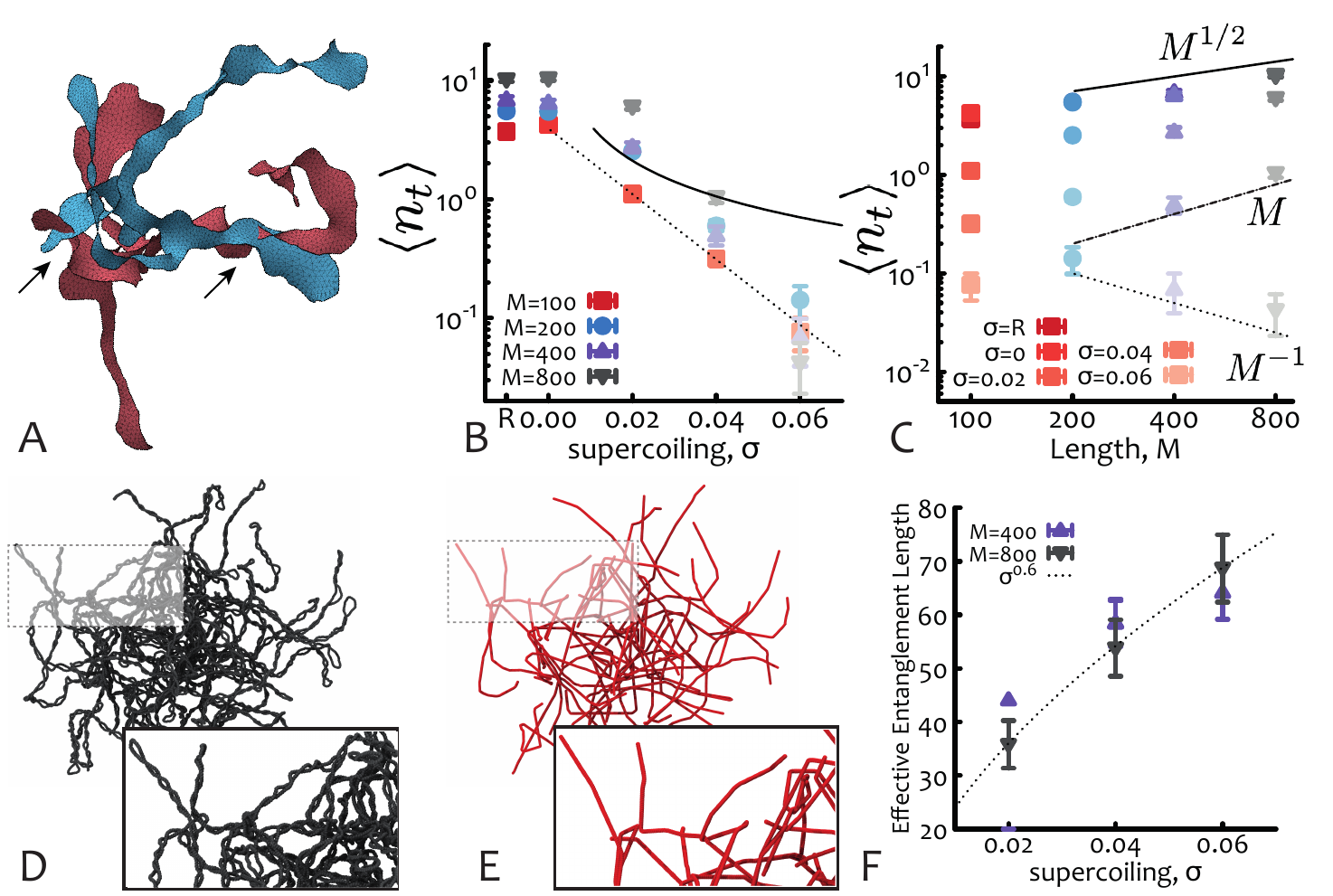}
	\vspace*{-0.4 cm}
	\caption{\textbf{Supercoiling Reduces Threadings and Entanglements.} \textbf{A} Snapshot of two threading plasmids (relaxed, $M=800$) with minimal surfaces drawn and intersections highlighted by arrows. \textbf{B} Number of threadings per plasmid as a function of supercoiling (dashed=exponential, solid=$1/\sigma$). \textbf{C} Number of threadings per plasmid as a function of plasmid length (dashed=$1/M$, solid=$M^{1/2}$, dot-dashed=$M$). \textbf{D,E} Snapshots of the PPA analysis run on our system with plasmids $M=800$ and $\sigma=0.06$. \textbf{F} The effective entanglement length increases with supercoiling as $N_e \sim s^{0.6}$ for $M=800$. }
	\label{fig:Threading_Entanglement}
	\vspace*{-0.6 cm}
\end{figure*}

Motivated by the observation that the minimal surface -- or ``threadable area'' -- decreases with supercoiling, we decided to more precisely quantify the number of threadings per plasmid for different levels of supercoiling. 
To this end we identify a plasmid to be ``passively threaded'' by another when the minimal surface of the former is intersected by the contour of the latter (at least twice, as they are topologically unlinked)~\cite{Smrek2016} (Fig.~\ref{fig:Threading_Entanglement}A).  As shown in Fig.~\ref{fig:Threading_Entanglement}B, the average number of threadings per plasmid $\langle n_t \rangle$ appears to decrease exponentially with supercoiling and to mirror the behaviour of $\langle \Sigma \rangle$. [As for the minimal surface, a notable exception to this general trend is the case of short plasmids ($M=100$) for which we find that $\langle n_t \rangle$ is statistically larger for $\sigma=0$ than for relaxed plasmids.] 

Based on these findings, we can also advance an argument as for why the mobility of the plasmids should increase with supercoiling: it was recently found that the dynamics of polymeric systems populated by threading topological constraints -- for instance melts of tadpole-shaped polymers~\cite{Rosa2020,Doi2015a} or compressed long plasmids~\cite{Soh2019}  -- slows down exponentially (with the number of threadings~\cite{Rosa2020}). As seen in this section, increasing supercoiling reduces the threadable area and, in turn, the actual number of threadings; we thus expect the dynamics of  highly supercoiled (threading-poor) plasmids to be exponentially faster than their relaxed (threading-rich) counterparts as seen in Fig.~\ref{fig:TAMSD}C. In spite of this our results are not definitive, and more work ought to be done. 
\vspace{-0.2 cm}

\subsection*{Supercoiling Reduces Entanglements}

The shape descriptors studied above suggest that the plasmids assume prolate double-folded conformations but it remains unclear whether the conformations are simply plectonemic (linear-like) or more branched into comb, star or tree-like structures~\cite{Everaers2017}. We thus computed the local absolute writhe along the contour length, $W(s)$, from which the number and location of plectonemic tips can be extracted as the maxima of this function~\cite{Michieletto2016softmatter,Vologodskii_etal1992} (see Methods). This calculation reveals that most of the conformations for $\sigma \geq 0.04$ have 2 tips consistent with \cite{Vologodskii_etal1992}, i.e. they mainly assume linear-like plectonemic conformations for the lengths considered in this work (see Fig.~S5 in SI). [For smaller supercoiling it cannot unambiguously distinguish tips from other regions of large curvature.]

In light of this finding another apparent controversy arises. Indeed, arguably, linear chains of half the length as their ring counterparts diffuse slower than the rings due to reptation relaxation induced by ordinary entanglements of linear chains~\cite{Halverson2011dynamics}; instead, we observe the opposite trend. To explain this we adapted the primitive path analysis (PPA) method~\cite{Everaers2004} (see Fig.~\ref{fig:Threading_Entanglement}C,D and Methods) and determined an effective entanglement length $N_{\textrm{e}}$ for highly supercoiled plasmids leveraging the fact that the tips of linear-like or branched conformations represent effective termini that can be pinned down. We find that $N_{\textrm{e}}$ grows with $\sigma$ (Fig.~\ref{fig:Threading_Entanglement}F), suggesting that the larger the supercoiling the less entangled the plasmids. For instance, we find that the effective linear backbone of the longest relaxed plasmids ($M=400$) with $\sigma=0.06$ is only about 3 entanglement lengths long. We argue that this unexpected supercoiling-driven reduction in entanglements is due to the fact that supercoiling increases the local concentration of intra-chain beads~\cite{Vologodskii1992} in turn inflating the effective tube surrounding any one plasmid. Notice that for short plasmids the PPA method cannot identify entanglement length, confirming that these are very poorly entangled and hence only weakly affected by inter-chain entanglements (thus confirming our argument for the behaviour of $R_g$ in Fig.~\ref{fig:RG}).

\vspace{-0.2 cm}
 \paragraph{Conclusions } 
 
 In this work we have simulated entangled solutions of DNA plasmids to study how supercoiling can be leveraged to tune the DNA dynamics orthogonally to other traditional methods, such as length or concentration. We have discovered that, contrarily to what typically assumed, the size of long plasmids increases with supercoiling when in entangled solutions. Surprisingly, we find that instead the mobility of the plasmids is enhanced by supercoiling. We discovered that this is due to severely asymmetric conformations which greatly reduce the number of threadings. In parallel, also generic entanglements are reduced as supercoiling increases the local concentration of intra-chain contacts thus effectively inflating the tube formed by neighbouring chains. We thus discovered that the unexpected enhanced diffusivity of entangled supercoiled DNA is due to a combination of reduced threading and entanglements.
 
We conjecture that highly supercoiled, longer plasmids in entangled solutions may display branched architectures, triggering the need of arm retraction or plectoneme diffusion/hopping relaxation mechanisms; in turn, this would entail a re-entrant slowing down of the plasmids diffusion. This non-monotonic dependence of DNA mobility on supercoiling would allow even richer control of the rheological properties.
 

Given our highly counter-intuitive results and the marked effect of supercoiling on plasmids microrhelogy we believe that our work will stimulate in vitro experiments using DNA plasmids extracted from bacteria and screened for desired levels of supercoiling. Albeit difficult, this is now nearly technically feasible~\cite{Peddireddy2019}. Ultimately, understanding how DNA topology and supercoiling affect the dynamics and conformational properties of plasmids in \emph{entangled} or \emph{crowded} conditions may not only reveal novel pathways to achieve fine tuning of the rheology of biopolymer complex fluids but also shed light on fundamental biological processes.

\section*{Materials and Methods}
\subsection*{Molecular Dynamics}
	Each bead in our simulation is evolved through the Langevin equation
	$m_a \partial_{tt} \vec{r}_a  = - \nabla U_a - \gamma_a \partial_t \vec{r}_a + \sqrt{2k_BT\gamma_a}\vec{\eta}_a(t)$, where $m_a$ and $\gamma_a$ are the mass and the friction coefficient of bead $a$, and $\vec{\eta}_a$ is its stochastic noise vector satisfying the fluctuation-dissipation theorem. $U$ is the sum of the energy fields (see SI). The simulations are performed in LAMMPS~\cite{Plimpton1995} with $m = \gamma = k_B = T = 1$ and using a velocity-Verlet algorithm with integration time step $\Delta t = 0.002\,\tau_{B}$, where $\tau_{B} = \gamma \sigma^2/k_BT$ is the Brownian time. 
	
\vspace{-0.3 cm}
	\subsection*{Branching Analysis}
Following Refs.~\cite{Michieletto2016softmatter,Klenin2000}, we compute the absolute writhe of a segment of a plasmid as $W(s)=(1/4\pi) \int^s_{s-l} \int_s^{s+l} |(\bm{r}_1-\bm{r}_2) \cdot (d\bm{r}_1 \times d\bm{r}_2)/|\bm{r}_1 - \bm{r}_2|^3 | $ with window $l=50$ beads. This calculation yields a function $W(s)$ whose maxima represent regions of high local writhe and can identify tips of plectonemes. In addition to being a local maximum, we require that $W(s)>0.35$ to avoid false positives. See SI for more details. 
	
\vspace{-0.3 cm}
	\subsection*{Primitive Path Analysis}
Following Ref.~\cite{Everaers2004}, we fix certain polymer segments in space, turn intra-chain repulsive interactions off and keep inter-chain interactions on. We then run simulations at low temperature $0.01$ to find a ground state. The resulting chain conformations (primitive paths) are made of straight segments connected by sharp kinks due to entanglements. The entanglement length is then given by $N_{e} = r_{\textrm{ee}}^{2}/(M b_{pp}^{2})$, where $r_{\textrm{ee}}$ is the mean endpoint distance, $M$ is the number of monomers between the fixed points and $b_{pp}$ is the mean bond-length of the primitive path. We adapt the classical PPA for plasmids by fixing the tips of all detected plectonemes instead of the end points of linear chains (see SI).

\subsection*{Acknowledgements}
The authors would like to acknowledge the networking support by the ``European Topology Interdisciplinary Action'' (EUTOPIA) CA17139. This project has received funding from the European Union’s Horizon 2020 programme under grant agreement No. 731019 (EUSMI). DM acknowledges the computing time provided on the supercomputer JURECA at J\"ulich Supercomputing Centre and the Leverhulme Trust (ECF-2019-088). JS acknowledges the support from the Austrian Science Fund (FWF) through the Lise-Meitner Fellowship No.~M 2470-N28. JS is grateful for the computational time at Vienna Scientific Cluster. Sample codes can be found at \url{git.ecdf.ed.ac.uk/dmichiel/supercoiledplasmids}.


\bibliography{library}

\end{document}